\documentclass[preprint,
superscriptaddress,
showpacs,
 amsmath,amssymb,
]{revtex4-1}
\usepackage{graphicx}
\usepackage{dcolumn}
\usepackage{bm}
\usepackage{color}
\usepackage{ulem}

\newcommand{\nab}{{\bf{\nabla}}}

\newcommand{\pa}{\partial}

\newcommand{\largep}{{\bf p}}

\newcommand{\largeA}{{\bf A}}

\newcommand{\ep}{\varepsilon}

\newcommand{\larger}{{\bf r}}

\begin{document}

\title{Fast-forward scaling theory}

\author{
Shumpei Masuda}
\affiliation{Research Center for Emerging Computing Technologies, National Institute of Advanced Industrial Science and Technology (AIST), 1-1-1, Umezono, Tsukuba, Ibaraki 305-8568, Japan}
\email{shumpei.masuda@aist.go.jp}

\author{
Katsuhiro Nakamura}
\affiliation{Faculty of Physics, National University of Uzbekistan, Vuzgorodok, Tashkent 100174, Uzbekistan}




\begin{abstract}
Speed is the key to further advances in technology.
For example, quantum technologies, such as quantum computing, require fast manipulations of quantum systems in order to overcome the effect of decoherence.
However, controlling the speed of quantum dynamics is often very difficult due to both the lack of a simple scaling property in the dynamics and the infinitely large parameter space to be explored.
Therefore, protocols for speed control based on understanding on the dynamical properties of the system, such as non-trivial scaling property, are highly desirable.
Fast-forward scaling theory (FFST) was originally developed to provide a way to accelerate, decelerate, stop and reverse the dynamics of quantum systems. 
FFST has been extended in order to accelerate quantum and classical adiabatic dynamics of various systems including cold atoms, internal state of molecules, spins and solid-state artificial atoms.
This paper describes the basic concept of FFST and review the recent developments and its applications such as fast state-preparations, state protection and ion sorting.
We introduce a method, called inter-trajectory travel, derived from FFST recently. 
We also point out the significance of deceleration in quantum technology.
\end{abstract}








\maketitle

\section{Introduction}
Speed is of prime significance for humans and probably for all living things.
The pursuit of speed in our life, strongly driven by economical competitions, has led to the development of technology in various fields including transportation, agriculture, communication, business and industry. 
Quantum technologies, which utilize the quantum nature of microscopic objects, have been growing rapidly after the discovery of the quantum world in last century.
Unfortunately, we need to be even faster in the quantum world.
The significance of speed in the current quantum technology is rooted not only in economical competitions but also (or more directly) in the short life time of a quantum state, which is shorten by the decoherence due to an undesired interaction with an environment.
We need to perform controls of a quantum system in a shorter time than its coherence time in order to use its quantum nature.
Therefore, fast and accurate control is an essential ability for further development of quantum technologies.

Current technology allows one to tune an external field to control a quantum system, in a period of time shorter than the coherence time of the system.
However, in general, it is very difficult to find the time dependence of the external fields (we refer control parameters hereafter), which can generate a target state with high accuracy at a predetermined short time, due to both the lack of a simple scaling property in the dynamics and the infinitely large parameter space to be explored.
Naive numerical trial-and-error simulations of the system to find a proper time dependence of control parameters are too time- and resource-consuming, when the Hilbert space of the system is large. 
Optimization of control parameters for desired controls is not an insignificant technical problem of individual laboratories, but an essential and omnipresent problem inherent in quantum technology that should be addressed systematically by the discipline of control of physical systems.
Systematic methods for fast and accurate controls of quantum systems based on understanding on their dynamical properties are highly desirable.

Is it possible to control the speed of quantum dynamics?
In 2008, this fundamental question was answered affirmatively by Masuda and Nakamura~\cite{Masuda2008,Masuda_thesis,Masuda2016}.
They introduced the fast-forward scaling theory (FFST), which provides a way to find control parameters which can accelerate, decelerate, stop or reverse a given reference dynamics of a quantum system.
This speed-control protocol is highly desirable for quantum technologies in terms of state preparation.
`To find a control parameter which generates a target state at time $T$ from a given initial state' is an ubiquitous task in quantum technologies.
FFST can considerably reduce the difficulty of the control problem to `find a control parameter which generates a target state from a given initial state' removing the constraint on time~\cite{Masuda2018}.

\begin{figure}[h!]
\includegraphics[width=3.8in]{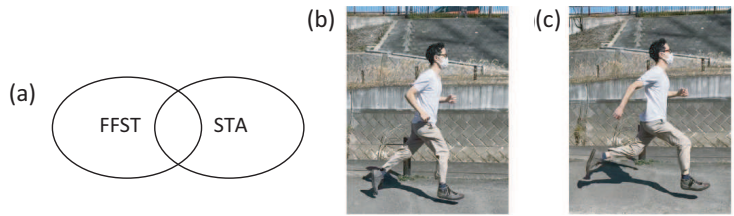}
\caption{
(a) FFST and STA have overlap. FFST for quantum adiabatic dynamics can be categorized to STA.
(b,c) Running forms of one of the authors (S.M.), jogging (b) and sprinting (c). }
\label{running2_4_6_22}
\end{figure}

The original FFST was developed for speed control of non-adiabatic dynamics of a single particle, Bose-Einstein condensate (BEC)~\cite{Masuda2008,Masuda_thesis} and two level systems~\cite{Masuda_thesis}.
In 2009, FFST for quantum adiabatic dynamics was developed, for a single particle and a BEC, to generate the end state of a quantum adiabatic dynamics in short time~\cite{Masuda2010}. 
The acceleration of adiabatic dynamics based on FFST can be categorized to what's known as shortcuts to adiabaticity (STA)~\cite{Torrontegui2013,Masuda2016,Campo2019,Guery-Odelin2019}, a group of protocols, which mitigate or eliminate completely unwanted energy excitations and realize the desired end state of adiabatic dynamics in short time.
Thus, the notions of FFST and STA have overlap as depicted in Fig.~\ref{running2_4_6_22}(a).
Then, FFST was extended with great effect to charged particles~\cite{Masuda2011,Kiely2015,Setiawan2017,Setiawan2019c},
many-body~\cite{Masuda2012} and discrete systems~\cite{Masuda2014,Takahashi2014,Xi2021}, tunneling ~\cite{Khujakulov2016,Nakamura2017} and Dirac dynamics \cite{Deffner2016,Sugihakim2021}. 
Acceleration of classical adiabatic dynamics~\cite{Jarzynski2017}, role of FFST in non-equilibrium statistical mechanics~\cite{Babajanova2018,Nakamura2020} and various controls of cold atoms~\cite{Masuda2010,Torrontegui2012,Masuda2014b,Martinez-Garaot2016} have been the subjects of researches.
An approach encompassing quantum, classical and stochastic dynamics~\cite{Patra2017} and 
a semiclassical approach~\cite{Patra2021} were proposed.

This paper describes the basic concept of FFST and review its applications.
While there is another excellent review focusing on FFST and counter-diabatic protocol~\cite{Masuda2016},
this paper covers more recent developments in the framework and applications including a novel method, called inter-trajectory travel, derived from FFST. 
In this paper, we also emphasize that FFST can be used for deceleration, stopping and reversing quantum dynamics as well as acceleration, and point out the significance of deceleration in quantum technology.
Discipline of control of physical systems enjoy a diversity of protocols, including optimal control theory, STA and FFST. 
The diversity is beneficial in applications, since each system to be controlled has different characteristics and experimental constraints.
In this field, each control protocol is sometimes viewed as a tool and the discipline as a toolbox which lead to technological developments.
Therefore, a more versatile and diverse range of tools should be further enhanced.
Further great advance in this field will be brought, in future, as a development of a platform where people can easily find optimal control parameters without selecting control protocols nor even recognizing which protocols were used.
Such platform might be an artificial intelligence which selects and combines proper control protocols and provides desired parameters, with an input of a little information such as Hamiltonian, experimentally feasible parameter range or even a measured data set. 
In this sense, the ultimate goal of study of controls is to be forgotten by people, although it will take more time. 
We hope that this review paper on FFST will remind readers of what is now in our toolbox, help in the development of the platform, and entertain those interested in non-trivial scaling properties in the dynamics of physical systems.

\section{Basic concept}
\label{Basic concept}
An important procedure in FFST is to look for a viable state of the system during the speed control, taking into account physical constraints.
This resembles the fact that our running form changes with speed. 
As you may notice, our body is not simply moving faster when we run faster (see Fig.~\ref{running2_4_6_22}(b)).
It is because that we are in the air for a certain period of time after kicking off the ground, and fall at a constant acceleration due to the gravity no matter what speed we are moving horizontally.
Thus, simply accelerated motion can not be realized due to the fixed gravity. 
In this section, we illustrate the basic concept of FFST showing how to find the proper state of a quantum system in the speed control~\cite{Masuda2008}. 

\subsection{Speed control of quantum dynamics of a single particle}
\begin{figure}[h!]
\centering\includegraphics[width=2.5in]{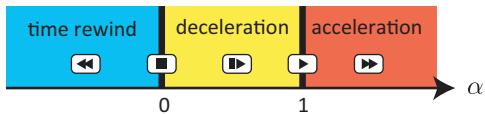}
\caption{
Illustration of the role of the speed-control factor $\alpha$. The speed-control factor characterizes the degree of speed control and is time dependent. The original dynamics is accelerated for $\alpha(t) > 1$, slowed for $0 < \alpha(t) < 1$, driven backward for $\alpha(t) < 0$, and stopped for $\alpha(t)=0$. $|\alpha|\gg 1$ can be used for STA.
}
\label{fig_sim}
\end{figure}

As an example, we consider speed control of a quantum dynamics of a single particle in potential $V_0(\larger,t)$.  
A reference dynamics, of which speed is controlled, is defined by the wave function $\Psi_0(\larger,t)$ of the particle, and $\Psi_0(\larger,0)$ and $\Psi_0(\larger,T)$ are referred as the initial and target (final) state, respectively.
The wave function satisfies the Schr\"odinger equation 
\begin{eqnarray}
i\hbar\frac{\pa\Psi_0(\larger,t)}{\pa t} = - \frac{\hbar^2}{2m}\nab^2\Psi_0(\larger,t) + V_0(\larger,t)\Psi_0(\larger,t),
\end{eqnarray}
where $m$ is the mass of the particle.
The wave function of a simply speed-controlled dynamics can be written as $\Psi_0(\larger,\Lambda(t))$ with 
\begin{eqnarray}
\Lambda(t)=\int_0^tdt'\alpha(t'),
\label{Lambda_3_3_22}
\end{eqnarray}
where the speed-control factor $\alpha(t)\in R$ (called magnification factor in Ref.~\cite{Masuda2008}) characterizes the degree of speed control. 
The reference dynamics is accelerated for $\alpha(t) > 1$, slowed for $0 < \alpha(t) < 1$, driven backward for $\alpha(t) < 0$, and stopped for $\alpha(t)=0$ (see Fig.~\ref{fig_sim}). 
For example, the speed-controlled dynamics evolves twice faster than the reference one for $\alpha=2$; the speed-controlled dynamics is time-reversal for $\alpha=-1$.
If we can realize the simply speed-controlled dynamics, we can create the target state at a desired time $T_{\rm F}(\ne T)$, where $T_{\rm F}$ is related to $T$ by $\Lambda(T_{\rm F}) = T$.
However, the simply speed-controlled dynamics requires an imaginary potential or modulation of the mass, which are not realistic usually~\cite{Masuda2008}.

An essential procedure in FFST is to modulate the wave function of the speed-controlled dynamics, so that the potential is to be real and the mass is to be fixed. 
Masuda and Nakamura found a proper wave function of the speed-controlled dynamics, written as~\cite{Masuda2008}
\begin{eqnarray}
\Psi_{\rm FF}(\larger,t) = \Psi_0(\larger,\Lambda(t))\exp[if(\larger,t)].
\label{PsiFF1}
\end{eqnarray}
This modified speed-controlled wave function (hereafter we refer to speed-controlled wave function) has an additional phase $f(\larger,t)$ compared to the simply speed-controlled one, while the amplitude is the same.
The driving potential $V_{\rm FF}$, which generates $\Psi_{\rm FF}(\larger,t)$, can be explicitly written as~\cite{Masuda2008}
\begin{eqnarray}
V_{\rm FF}(\larger,t) &=& -\hbar\frac{d\alpha(t)}{dt}\eta(\larger,\Lambda(t)) 
-\hbar \frac{\alpha^2(t)-1}{\alpha(t)}\frac{d\eta(\larger,\Lambda(t))}{dt}\nonumber\\
&&-\frac{\hbar^2}{2m}[\alpha^2(t)-1] [\nab\eta(\larger,\Lambda(t))]^2 + V_0(\larger,\Lambda(t)).
\label{VFF1}
\end{eqnarray}
In Eq.~(\ref{VFF1}), $\eta(\larger,t)$ is the phase of the reference wave function, that is, the reference wave function can be written as $\Psi_0(\larger,t) = \tilde{\Psi}_0(\larger,t)\exp[i\eta(\larger,t)]$ with $\tilde{\Psi}_0(\larger,t)$, $\eta(\larger,t)\in R$, where $\tilde{\Psi}_0(\larger,t)$ is the amplitude of the reference wave function.
The additional phase is explicitly written as 
\begin{eqnarray}
f(\larger,t) = [\alpha(t)-1]\eta(\larger,\Lambda(t)).
\label{f_1}
\end{eqnarray}
The last term in Eq.~(\ref{VFF1}) is the original potential with scaled time $\Lambda(t)$. 
Equation~(\ref{VFF1}) indicates that the auxiliary potential, $V_{\rm FF}(\larger,t)-V_0(\larger,\Lambda(t))$, should be added to the original one in order to generate the speed-controlled dynamics.
The auxiliary potential tends to become large with the increase of $|\alpha(t)|$.

The most important  and surprising consequence of FFST is that the additional phase can be made to vanish at any desired time $t'$ by choosing the speed-control factor to satisfy $\alpha(t') = 1$, as clearly seen in Eq.~(\ref{f_1}). 
Thanks to this useful property, we can generate the exact target state at arbitrary $T_{\rm F}$ from the initial state by choosing the time-dependence of the speed-control factor to satisfy $\alpha(0) = \alpha(T_{\rm F}) = 1$.
Here, $T_{\rm F}$ can be arbitrarily short as long as the strong auxiliary potential does not change the internal state of the particle or relativistic effects do not become significant.  

\textit{\textbf{Scaling property---}} 
FFST reveals non-trivial scaling property of quantum dynamics, which connects infinite number of different dynamics with different speed. 
More explicitly, once we have a pair composed of a viable wave function and the corresponding potential, 
$\{ \Psi_0(\larger,t),V_0(\larger,t)\}$, FFST presents a way to obtain the speed controlled pair, $\{ \Psi_{\rm FF},V_{\rm FF}\}[\alpha]$, for arbitrary $\alpha(t)$ which has a finite time derivative.
The original pair can be represented as $\{ \Psi_{\rm FF},V_{\rm FF}\}[1]$.

\subsection{Deceleration}
\label{Deceleration}
Our lives demand greater speed in order to compete with others and to adapt to new technologies and cultures.
Admittedly benefiting from speed, we are sometimes fed up with the pressure to the acceleration, and desire slowing down our lives and freeing ourselves from the obsession with speed. 
Ironically, however, what slowdown of quantum dynamics will bring is the further acceleration of quantum technology.

There are always hardware limitations to the rate at which control parameters can be changed.
Suppose that a laboratory (Lab A) has a hardware, which can generate a rapidly changing control field, and they kindly share the information of the time dependence of the control field that can produce nice results; on the other hand, another laboratory (Lab B) has only a hardware, which can generate slower control field. Do we need another time- and resource-consuming simulations to optimize the control parameters for the slower hardware of Lab B? With this question in mind, deceleration of quantum dynamics was investigated, and they showed the way to find control parameters varying more slowly and still can generate approximately the same end state as the original control of Lab A~\cite{Masuda2021}.
This technique can be useful also in the sense that quantum system is typically made to interact with other quantum objects or classical hardwares in quantum technologies. 
Sometimes we need to make the quantum system to reach to a desired state in later time to wait other hardwares or other quantum systems.
Thus, the deceleration is useful as well as acceleration to change the timing when quantum state reach to a target state.

\section{Acceleration of quantum adiabatic dynamics}
Adiabatic control, which has recourse to quantum adiabatic dynamics, enjoys broad applicability in physics and chemistry. 
However, the efficiency of adiabatic control, in which control parameters are changed slowly to avoid non-adiabatic transitions, is degraded by decoherence. 
Acceleration of quantum adiabatic dynamics to overcome the effect of decoherence will find wide applications in various area.
FFST was extended in order to accelerate quantum adiabatic dynamics and generate its end state in a short period of time~\cite{Masuda2010}.
In this section, we explain the basic concept of FFST for the quantum adiabatic dynamics.


As an example, we illustrate the acceleration of a quantum adiabatic dynamics of a particle in potential $V_0 = V_0(\larger, R_a(t))$~\cite{Masuda2010}, where
$R_a(t) = R_i + \ep t$ is a parameter that determines the potential profile. 
We assume that the initial state is in the $n$th energy eigenstate and the
rate of change of $R_a(t)$ is infinitesimal, i.e., $\ep \ll 1$. 
Quantum adiabatic theorem states that the system remains in the $n$th energy eigenstate of the instantaneous Hamiltonian.
In the ideal adiabatic dynamics, the wave function can be written as 
\begin{eqnarray}
\Psi_0(\larger,t,R_a(t)) = \phi_n(\larger,R_a(t))\exp\Big{[}-\frac{i}{\hbar}\int_0^tE_n(R_a(t'))dt'\Big{]},
\label{Psi0_2_28_22}
\end{eqnarray}
where $\phi_n(\larger,R_a)$ and $E_n(R_a)$ are the wave function and the energy of the $n$th eigenstate of the instantaneous Hamiltonian $H = \largep^2/(2m)+V_0(\larger, R_a)$.
In Eq. (\ref{Psi0_2_28_22}), we consider the case that there is no adiabatic phase for simplicity.
We aim to drive the system from $\phi_n(\larger,R_a(0))$ to $\phi_n(\larger,R_a(T))$ in a short time, while the original final time $T$ is of $O(\varepsilon^{-1})$.
In FFST for the adiabatic dynamics, we accelerate the infinitely slow dynamics with the infinitely large speed-control factor.
More explicitly, we consider the limit $\varepsilon\rightarrow 0$, and we put the speed control factor $\alpha$ to be $O(\varepsilon^{-1})$ so that $\alpha \varepsilon\sim 1$.

In order to apply FFST developed in the previous section, 
a set of the wave function and potential, which satisfies the Schr\"{o}dinger equation up to $O(\varepsilon)$, is introduced. This process is called regularization.
The regularized wave function and potential are written as~\cite{Masuda2010}
\begin{eqnarray}
\Psi_0^{(\rm reg)}(\larger,t) &=& \phi_n(\larger,R_a(t))\exp\Big{[}-\frac{i}{\hbar}\int_0^tE_n(R_a(t'))dt'\Big{]}\exp[i\ep\theta(\larger,R_a(t))],\nonumber\\
V_0^{(\rm reg)}(\larger,t) &=& V_0(\larger,R_a(t)) + \varepsilon \tilde{V}(\larger,R_a(t)),
\label{Psi0_1}
\end{eqnarray}
where
\begin{eqnarray}
\tilde{V}(\larger,R_a) = -\hbar\mbox{Im} \Big{[}\frac{1}{\phi_n}\frac{\pa\phi_n}{\pa R_a}\Big{]}
-\frac{\hbar^2}{m}\mbox{Im}\Big{[}\frac{\nab\phi_n}{\phi_n}\Big{]}\cdot\nab\theta,
\end{eqnarray}
and $\theta$ satisfies
 \begin{eqnarray}
 |\phi_n|^2\nab^2\theta + 2\mbox{Re}[\phi_n\nab\phi_n^\ast]\cdot\nab\theta
 +\frac{2m}{\hbar}\mbox{Re}\Big{[}\phi_n\frac{\pa\phi^\ast_n}{\pa R_a}\Big{]}=0.
 \end{eqnarray}
We apply FFST introduced in the previous section to this regularized adiabatic dynamics.
The wave function of the accelerated state and driving potential are obtained as 
\begin{eqnarray}
\Psi_{\rm FF}(\larger,t) &=& e^{if(x,t)}\Psi_{0}^{(\rm reg)}(x,\Lambda(t)),\label{PsiFF_3_3_22}\\
V_{\rm FF}(\larger,t) &=& V_0(\larger,\Lambda(t)) + \alpha(t)\varepsilon \tilde{V}(\larger,\Lambda(t)) -
\hbar\frac{d\alpha(t)}{dt}\varepsilon\theta(\larger,\Lambda(t))
-\hbar\alpha^2(t) \varepsilon^2\frac{\partial\theta(\larger,R_a(\Lambda(t)))}{\partial R_a}\nonumber\\
&&  - \frac{\hbar^2}{2m_0}\alpha^2(t)\varepsilon^2 [\nabla\theta(\larger,R_a(\Lambda(t)))]^2,
\label{VFF_3_3_22}
\end{eqnarray}
where $\Lambda(t)$ is defined by Eq.~(\ref{Lambda_3_3_22}), and $f(\larger,\Lambda(t))=\alpha\varepsilon \theta(\larger,R_a(\Lambda(t)))$.

In order to make the above results more usable, we can alternatively use the wave function written as
\begin{eqnarray}
\Psi_{\rm FF}(\larger,t) = \phi_n(\larger,R(t))\exp\Big{[}-\frac{i}{\hbar}\int_0^tE_n(R(t'))dt'\Big{]}\exp[if(\larger,t)],
\label{PsiFF_2}
\end{eqnarray}
which is identical to Eq.~(\ref{PsiFF_3_3_22}) except for the space-independent phase.
Here, $R(t)$ is defined by $R(t)=R_a(\Lambda(t))$, and changes non-adiabatically in contrast to $R_a(t)$.
The driving potential is represented as
\begin{eqnarray}
V_{\rm FF}(\larger,t) &=& V_0(\larger,R(t)) - \frac{\hbar^2}{2m}[\nab f(\larger,t)]^2 - \frac{\hbar^2}{m}\nab f(\larger,t) \cdot \mbox{Im}
 \Big{[} \frac{\nab\phi_n(\larger,R(t))}{\phi_n(\larger,R(t))} \Big{]}\nonumber\\
 &&-\hbar\frac{dR(t)}{dt}\mbox{Im}\Big{[} \frac{1}{\phi_n(\larger,R(t))} \frac{\pa \phi_n(\larger,R(t))}{\pa R}\Big{]}
 -\hbar\frac{df(\larger,t)}{dt}.
 \label{VFF_3_4_22}
 \end{eqnarray}
 The additional phase, which is alternatively written as $f(\larger,t)=(dR/dt)\theta$, can be obtained by solving
 \begin{eqnarray}
 |\phi_n(\larger,R(t))|^2\nab^2f(\larger,t) + 2\nab f(\larger,t)\cdot \mbox{Re} [\phi^\ast_n(\larger,R(t)) \nab\phi_n(\larger,R(t))]\nonumber\\
 +\frac{2m}{\hbar}\frac{dR}{dt}\mbox{Re} \Big{[} \phi_n^\ast(\larger,R(t)) \frac{\pa \phi_n(\larger,R(t))}{\pa R}\Big{]} = 0
 \label{eqf_1}
 \end{eqnarray}
subject to the boundary conditions  $R(0)=R_i$, $R(T_{\rm F})=R_f$ and $dR/dt(0)=dR/dt(T_{\rm F})=0$. The state of the system is driven from $\phi_n(\larger,R_i)$  to $\phi_n(\larger,R_f)\exp\Big{[} -(i/\hbar)\int_0^{T_{\rm F}}E_n(R(t'))dt'\Big{]}$ in time $T_{\rm F}$ under the driving potential.   

Direct derivation (called streamlined version) of the additional phase and driving potential was shown in Ref.~\cite{Torrontegui2012,Torrontegui2013,Kiely2015,Martinez-Garaot2016} without considering the acceleration of a slow reference adiabatic process. 
Connection of the FFST for adiabatic dynamics to other STA protocols, such as invariant based~\cite{Torrontegui2011,Torrontegui2012} and counter-diabatic protocols~\cite{Takahashi2014,Takahashi2015,Jarzynski2017,Patra2017,Patra2021}, were also studied.

As seen in Eq.~(\ref{VFF_3_4_22}), the driving potential can diverge at the nodes of the wave function where $|\phi_n(\larger,R(t))|=0$. 
A simple, intuitive criterion for determining whether these singularities will arise, for a given excited state was developed~\cite{Patra2017}.
Moreover, a method which overcome the issue of the divergence was proposed~\cite{Patra2021}. 
 
\textit{\textbf{Scaling property---}} 
It is noteworthy that $f(\larger,t)(dR/dt)^{-1}$ is a function only of $\larger$ and $R$ as seen from Eq.~(\ref{eqf_1}). 
It implies that once $f(\larger,t)(dR/dt)^{-1}$ is obtained, the additional phase and driving potential for arbitrary time dependence of $R$ can be obtained without solving any equations.
In actual experiments, this scaling property is very useful because the duration of control may need to be changed for example when the equipments are replaced.
The scaling property considerably reduces time and effort to optimize the control parameter for each duration. 
There are always hardware limitations to the intensity of the driving potential.
The scaling property will allow one to satisfy such experimental constraint by modifying the time dependence of $R$, because the profile of the driving potential depends on the time dependence of $R$.


\subsection{Analytical formulae of driving potential}
The driving potential in Eq.~(\ref{VFF_3_4_22}) generally depends on the energy level $n$, and 
obtaining the set of the additional phase and driving potential requires numerical calculations.
However, the set of the additional phase and driving potential can be written analytically for 
some specific controls such as translation, compression/expansion and rotation.
Moreover, the auxiliary driving potential, $\Delta V_{\rm FF}(\larger,t)=V_{\rm FF}(\larger,t) - V_0(\larger,R(t))$, is independent of the reference potential $V_0(\larger,R(t))$.
For these reasons, the controls can be applied without detailed knowledge of the system such as profile of the original trapping potential and state of the system.
Here, we list the sets of  the additional phase and auxiliary driving potential.
\begin{table}[!h]
\caption{Analytical formulae of $\Delta V_{\rm FF}(\larger,t)$ are listed with relevant references for three different controls. Spatially uniform terms in $\Delta V_{\rm FF}(\larger,t)$ are omitted for simplicity. 
$m,q,$ and $c$ are the mass and charge of the particle and the speed of light, respectively. }
\label{table_example}
{\renewcommand\arraystretch{1.2}
\begin{tabular}{llll}
\hline
control & auxiliary driving potential $\Delta V_{\rm FF}$ & relevant references\\
\hline
translation & $-m\ddot{R}x$ & \cite{Masuda2010} \\
compression/expansion &  $( \frac{m\ddot{R}}{4R} - \frac{3m\dot{R}^2}{8R^2} )\larger^2$ & \cite{Masuda2010,Masuda2011} \\
rotation & $q^2/(2m c^2)[\largeA_0^2(\larger,R)-
 \largeA_{\rm FF}^2(\larger,t)]$  & \cite{Masuda2015a}\\
(vector potential $\Delta \largeA_{\rm FF}(\larger,t)$) & $(\dot{R}mc/q) [y,-x,0]$\\
\hline
\end{tabular}
}
\label{table_3_4_22}
\end{table}

\textit{\textbf{Translation---}}
We consider a translation of a particle in a stationary trapping potential $V(\larger)$ by modifying the potential.
The reference potential is given by $V_0(\larger,R)=V(x-R,y,z)$, where $R(t)$ is the translation distance.
The auxiliary driving potential for the translation~\cite{Masuda2010} is shown as Table~\ref{table_3_4_22}.
The additional phase is represented as $f(\larger,t) = \frac{m}{\hbar} \dot{R} x$, where dot denotes the time derivative.
This protocol was implemented for trapped ions~\cite{An2016}.
Translation of a charged particle in an electromagnetic field was studied in Ref.~\cite{Masuda2011}. 
It has also been shown that the driving potential for transport of particles interacting with each other is the same as the one particle system, and is independent of the detailed form of the interaction \cite{Masuda2012}.  
The reader is referred also to \cite{Chen2011, Torrontegui2011, Bowler2012, Torrontegui2012d} for investigations related to translation.
 
\textit{\textbf{Compression and expansion---}}
Compression/expansion of wave function was studied for a particle trapped by a harmonic  potential~\cite{Masuda2010} and for more general case with a scalar and vector potentials~\cite{Masuda2011}.
The reference potential for compression/expansion can be given by $V_0(\larger,R)=RV(\sqrt{R}\larger)$~\cite{Masuda2011}.
The auxiliary potential for compression/expansion~\cite{Masuda2011} is shown in Table~\ref{table_3_4_22}.
The additional phase is represented as $f(\larger,t) = -(m\dot{R}/4\hbar R)x^2$ 
(see Ref.~[\citenum{Masuda2011}] for more general case with a vector potential).
The reader is referred also to \cite{Muga2009, Muga2010, Schaff2010a, Schaff2011a, Schaff2011b, Chen2010a, Stefanatos2010, Stefanatos2011, Stefanatos2012, Campo2011, Campo2012a, Andresen2011, Torrontegui2012, Torrontegui2012c} for investigations related to compression/expansion.

\textit{\textbf{Rotation---}} 
Rotation of the wave function of a charged particle was studied~\cite{Masuda2015a}. The reference scalar and vector potentials for the rotation around the $z$-axis are represented as
\begin{eqnarray}
V_0(\larger,R) &=& V(x'(R),y'(R),z),\nonumber\\
\largeA_0(\larger,R) &=& \big{[}
\cos(R)A_x(\larger') - \sin(R)A_y(\larger'),\nonumber\\
&&\sin(R)A_x(\larger') + \cos(R)A_y(\larger'),
A_z(\larger')\big{]},
\label{VA}
\end{eqnarray}
where $\larger'=[x',y',z]$ and
$x'(R) = \cos(R)x+\sin(R)y, \ \ y'(R) = -\sin(R)x + \cos(R)y$.
The auxiliary scalar and vector potentials are shown in Table~\ref{table_3_4_22}.
We can set $f=0$ in this control (see Ref.~[\citenum{Masuda2015a}] for more general formula).

\textit{\textbf{State-dependent analytic solutions---}} 
State-dependent analytical formula of the additional phase and driving potential were reported.
In Ref.~\cite{Masuda2012}, the set of the additional phase and driving potential was shown analytically for  controls of the wave function with spherical symmetry.
Mart\'{i}nez-Garaot and co-authors derived the formal solution of the driving potential and additional phase
for the control during which the wave function is represented as $\Psi(x,t)=\rho(x,t)e^{i\phi(x,t)}$, where $\rho(x,t)$ is predetermined~\cite{Martinez-Garaot2016}.
It was also shown that FFST can be used to create the first excited state of a harmonic oscillator from the ground state.



\subsection{Many body system}
Theoretical framework of FFST was extended to control the speed of non-adiabatic dynamics and accelerate adiabatic dynamics of many-body systems~\cite{Masuda2012}. 
Although the driving potential is not viable in general because it includes many-body potentials, the driving potential is a one-body potential in specific cases. 
Importantly, the ideal transport, without excitation, of interacting identical particles can be realized by the one-body potential identical to the one for a single particle shown in Table~\ref{table_3_4_22}.
It was also shown that the driving potential becomes a one-body potential when the mean field approximation is valid for the energy eigenstates to be controlled.

\subsection{Discrete systems}
Population transfer in discrete systems or multi-level systems is ubiquitous phenomena in condensed matter physics including internal state of molecules, cold atoms in an optical trap and solid-state artificial atoms which can work as qubits. 
Therefore, speed control of the dynamics of discrete systems will find broad applications.
FFST for non-adiabatic dynamics of two-level systems was developed~\cite{Masuda_thesis}.
They introduced two types of FFST with different intermediate states.
One uses the additional phase of the intermediate state, and another uses more general unitary transformations.
The framework of FFST was extended to accelerate adiabatic dynamics of discrete systems~\cite{Masuda2014}.
In Ref.~\cite{Takahashi2014}, different approach which derives fast-forward Hamiltonian was introduced using a proper unitary transformation, and was applied to a two-level system.

Speed control of site-to-site population transfer of a BECs in an optical lattice was  
studied using FFST~\cite{Masuda2014}.
The theory was also applied to acceleration of stimulated Raman adiabatic passage (STIRAP) for selective vibrational population transfer in a polyatomic molecule~\cite{Masuda2015c}.
The motivation investigation was two-fold: (i) to overcome transfer inefficiency that occurs when the STIRAP fields and pulse durations must be restricted to avoid excitation of population transfers that compete with the targeted transfer and (ii) to overcome transfer inefficiency resulting from embedding of the actively driven subset of states in a large manifold of states.  
It was shown that, in a subset of states that is coupled to background states, a combination of STIRAP and fast-forward driving fields that do not individually generate processes that are competitive with the desired population transfer can generate greater population transfer efficiency than can ordinary STIRAP with similar field strength and/ or pulse duration. 
More recently, speed control of superconducting qubits, which are attracting attentions in terms of applications to quantum computing, was studied using FFST~\cite{Masuda2021}.

\subsection{Classical systems and semiclassical approach}
Classical counter part of FFST and semiclassical approach have been developed. 
In Ref.~\cite{Jarzynski2017}, the authors obtained the driving potential that guides classical trajectories started
from a given energy shell of an initial Hamiltonian to the corresponding energy shell of the final Hamiltonian, so that the initial and final values of action are identical for every trajectory.
The flow-fields approach which encompasses STA for quantum, classical and stochastic dynamics was developed in Ref.~\cite{Patra2017}. This approach can provide compact expressions for both the counter diabatic Hamiltonians and fast-forward driving potentials.
Semiclassical connection between quantum and classical STA was also investigated.
Remarkable advance in FFST was delivered by Patra and Jarzynski, which solves the issue of the divergence of the driving potential due to the nodes of the wave function of energy eigenstates~\cite{Patra2021}.
The authors developed fast-forward shortcuts free of the divergences.
Acceleration of adiabatic dynamics of the 17th energy eigenstate was numerically demonstrated.

\section{Applications}
The major application of FFST will be state preparation, aiming at creation of a target state at a predetermined time.
State preparation based on FFST was studied in various systems including BECs~\cite{Masuda2008,Masuda2010,Torrontegui2012,Masuda2014,Masuda2014b,Masuda2018,Xi2021}, interacting boson particles~\cite{Masuda2012}, charged particles~\cite{Masuda2011,Kiely2015,An2016,Khujakulov2016,Nakamura2017,Setiawan2021b},
vibrational levels of molecules~\cite{Masuda2015c}, spin systems~\cite{Takahashi2014,Setiawan2017,Setiawan2019c} and superconducting qubits~\cite{Masuda2021}.
Moreover, deceleration based on FFST will be useful for parameter optimization as explained in Sec.~\ref{Basic concept}\ref{Deceleration}.
In this section, we point out that FFST will find wider applicability such as protection of quantum state and ion sorting.

\subsection{Protection of quantum state from noise}
In actual experiments, a quantum system is subjected to noise. 
Therefore, development of methods for protecting quantum systems from the influence of noise is of practical importance. 
It was shown that sequential modulation of the state of systems can suppress the influence of decoherence caused by interaction with environment in two-level systems~\cite{Viola1998,Viola1999}. 
The technique is called dynamical decoupling.
Inspired by the technique, Masuda studied protection of quantum state from the influence of noise based on FFST~\cite{Masuda2013}.
FFST allows one to steer the state of systems continuously without undesired energy excitations, and the continuous modification of the state can cancel the influence of noise. 

\begin{figure}[!h]
\centering\includegraphics[width=12cm]{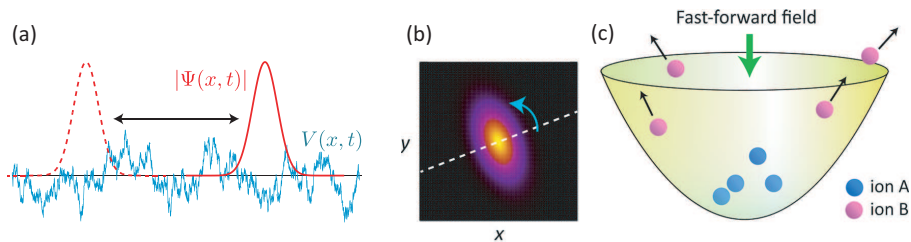}
\caption{(a) Schematic of successive transport of wave function under spatially random potential~\cite{Masuda2013}.
Schematic of rotation of the orientation of wave function distribution (b) and sorting of trapped ions with the fast-forward driving field (c)~\cite{Masuda2015a}.}
\label{protection_2_6_22}
\end{figure}

They considered a BEC, which is trapped in a harmonic potential and subjected to a fluctuating spatially random potential (see Fig.~\ref{protection_2_6_22}(a)).
It was shown that the alternating transport of the BEC based on FFST can suppress the influence of the random potential on the BEC. 
Note that, in this study, the successive transports do not aim at a transfer of the BEC to other place, but aim at keeping the same state as the initial state. 
Therefore, this application differs from many of other applications of STA to state preparation.

\subsection{Ion sorting}
Ion sorting based on FFST was proposed~\cite{Masuda2015a}.
Let us consider a mixture of different ionic species in a trap.
Sequential drive using FFST can retain a selected species un-excited in the trap, while the other species excited and eventually expelled from the trap.
This sorting scheme resembles sifting gemstones with a particular size from sands or, in sand painting, you shake a paper to remove unwanted sands while desired colored sands glued on the paper.
In Ref.~\cite{Masuda2015a}, transitionless rotation was utilized to 
remove unwanted ionic species from the trap while the wave function of target species is simply rotated 
around the trap center without excitation.
This scheme roots in the property of the driving fields: independent of the state of the target species and depending on the mass and charge of the target species.





\section{Role of fast-forward in non-equilibrium statistical mechanics}
In this section, we introduce the studies on non-equilibrium statistical mechanics of systems controlled based on FFST.
\subsection{Non-equilibrium equation of states of quantum gas in the fast forward of expanding cavity}
In the equilibrium equation of states  for an ideal classical gas (Boltzmann gas), the pressure ($P$), volume ($V$) and temperature ($T$) are quasi-static state variables and satisfy Boyle-Charles' law (BCL) and Poisson's adiabatic equation in the isothermal and thermally-adiabatic processes, respectively .  BCL is a special limit of the Bernoulli's formula  bridging between the pressure ($P$) and internal energy ($U$).  In the 1D (dimensional) cavity,  the force ($F$) and the length ($L$) play the role of
$P$ and $V$, respectively. The above equilibrium equation of states works as well in quantal gas systems. Here, the quasi-static process is equivalent to the quantal-adiabatic one which suppresses the transition among different quantal eigenstates.

How will the equilibrium equation of states be innovated when the motion of a quantal gas container would not be quasi-static?
In the case of a piston with a small but finite velocity, the non-adiabatic contribution proved not to affect the equilibrium equation of states seriously.
In the case of a rapid piston, however, we can expect a dramatic role of non-adiabatic contributions. 
The statistical treatment of a quantum gas is very difficult in general case of a rapid piston 
where the temporal change of state variables is far from being quasi-static. 
However, the fast forwarding of  adiabatic control of the confinement guarantees no transition among different quantum states, making such treatment feasible, and one can elucidate the exact relation among the rapidly-changing state variables.

Applying the idea of fast forward (FF) of adiabatic control of 1D confined systems, Babajanova, Matrasulov and Nakamura (BMN)~\cite{Babajanova2018} investigated the non-equilibrium equation of states of an ideal quantum gas (Fermi gas) confined to the rapidly dilating soft-wall and hard-wall cavities. 
BMN consider a Fermi gas of $N$ noninteracting particles confined in the harmonic potential whose frequency is time-dependent and also the gas system in the hard-wall confinement. BMN derived the non-equilibrium equation of states during the FF, and showed the fast-forward variants of Poisson's adiabatic equation and Bernoulli's formula.
Confining to the thermally-isolated isentropic process and using the exact solution of the von Neumann equation, statistical means of the quantal adiabatic and non-adiabatic forces are evaluated in both the low-temperature quantum-mechanical  and
high temperature quasi-classical regimes. Reflecting the fact that the fast-forward dynamics is population-preserving cooling or heating process, the state variables such as the statistical mean of force, cavity size and effective temperature are not quasi-static, but rapidly-changing variables.
BMN elucidated the quantum non-adiabatic (QNAD) contributions to Poisson's adiabatic equation
and to Bernoulli's formula.  The QNAD contributions are 
proportional to the acceleration of cavity size and square of its velocity, and has coefficients dependent on temperature, gas density and cavity size. BMN also revealed the condition when QNAD contribution overwhelms the quantal adiabatic one and throughly changes the feature of the conventional equilibrium equation of states.
On the other hand, BMN showed a scheme to realize the exact FF state with use of the electromagnetic field 
(see Sec. II.B of ~Ref.~[\citenum{Babajanova2018}]).
In that scheme $\Psi_{\rm FF}^{(0)}$ does not include the extra factor $e^{if}$ as in Eq.~(\ref{PsiFF_3_3_22}) of the present article and the FF protocol guarantees the perfect fidelity of the underlying adiabatic state throughout the time evolution until the completion of control protocol. Therefore the energetic cost caused by the FF driving is expected to be greatly decreased, which will be studied elsewhere by using the non-equilibrium equation of states.


\subsection{Fast forward of stochastic heat engine and its efficiency}
Carnot's concept of heat engines belongs to a classical subject of thermodynamics. To achieve the highest efficiency, a heat engine needs to operate a reversible thermodynamic cycle which requires  a quasi-static process and results in a vanishing power. The power means the work per one-cycle time. The quasi-static thermodynamic cycle should be speeded up so as to produce a finite power of realistic heat engines. 

In the context of nano-scale motors, Brownian heat engines have received a wide attention, which mimic a simple system of a stochastic heat engine whose degrees of freedom are subject to a time-dependent potential and working between hot and cold heat baths. The investigation on the efficiency of the engines of this kind  at maximum power  assumed the time dependence of the effective temperature (e.g., variance of the particle position) during the isothermal process. More recent works  proposed the engineering swift equilibration and the shortcut to isothermality, but provided neither kinetics corresponding to the thermally-adiabatic process nor investigation on the power and efficiency of the heat engine.

Nakamura, Matrasulov and Izumida (NMI) \cite{Nakamura2020} applied the FF scheme to the  classical stochastic Carnot-like heat engine which is driven by a Brownian particle coupled with a time-dependent harmonic potential and working between the high ($T_h$)- and low ($T_c$)-temperature heat reservoirs. The analysis is straightforward with use of the Kramers equation which works throughout in the full cycle of the heat engine.
Concentrating on  the underdamped case where momentum degree of freedom is  included, NMI found the explicit expressions for the FF protocols necessary to accelerate both the isothermal and thermally-adiabatic processes, and obtained the reversible and irreversible works.
The irreversible work is shown to consist of two terms with one proportional to and the other  inversely proportional to the friction coefficient. 
The optimal value of efficiency $\eta$ at the maximum power of this engine is found to be 
$\eta^*=\frac{1}{2} \left( 1+\frac{1}{2}\left(\frac{T_c}{T_h}\right)^{\frac{1}{2}} - \frac{5}{4}\frac{T_c}{T_h}  +O\left(\left(\frac{T_c}{T_h}\right)^{\frac{3}{2}}\right)\right)$ 
and $\eta^*= 1- \left(\frac{T_c}{T_h}\right)^{\frac{1}{2}}$, respectively in the cases of strong and weak dissipation.
The result is justified for a wide family of time scaling functions, making the FF protocols very flexible. 
NMI also revealed that the accelerated full cycle of the Carnot-like stochastic heat engine cannot be conceivable within the framework of the overdamped case, and the power and efficiency can be evaluated only when the momentum degree of freedom is taken into consideration. 



\section{ITT travel}
When utilizing FFST, one can often obtain viable trajectories in the state space, which realize the desired end state from a given initial state.
However, FFST is not applicable in some parameter regimes due to the lack of a viable fast-forward speed-controlled trajectories (FF trajectory), which connect the initial and target states. Therefore, modification of the theory is required to resolve this issue.
To this end, a novel scheme called inter-trajectory travel (ITT) was developed~\cite{Masuda2021}.

\begin{figure}[h!]
\centering\includegraphics[width=2.0in]{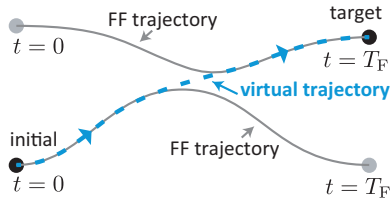}
\caption{
Illustration of inter-trajectory travel (ITT). A virtual trajectory interconnects two FF trajectories in order to connect the initial and target states.
}
\label{ITT_notion_3_7_22}
\end{figure}

We consider FF trajectories in the state space, which correspond to the speed controlled states realized by  FFST driving fields.
In some parameter regimes, there is no (or it is difficult to find) viable FF trajectory which connects the initial and target states.
In Ref.~\cite{Masuda2021}, the authors found two FF trajectories (see Fig.~\ref{ITT_notion_3_7_22}); one starts from the initial state; another ends up the target state at $t=T_{\rm F}$; and the two trajectory approach each other at a certain time.
They introduced a virtual trajectory interconnecting the FF trajectories, and derived the driving field which approximately realizes the virtual trajectory.
The theory was applied for acceleration, deceleration and STA of a two-qubit system.
The concept of ITT is anticipated to be used not only for FFST but also for other STA protocols.

\section{Summary}
We have reviewed the development and applications of FFST.
In order to overview the previous studies and illuminate directions for future research, we refer readers to Table~\ref{table_Classification_3_4_22} classifying FFST by systems and objectives etc~\cite{Masuda_thesis}.
Many of the previous studies of FFST would be classified according to the nature of the systems: (I) whether the dynamics is classical-mechanical or quantum-mechanical, (II) relativistic or non-relativistic, (III) whether we deal with a single particle/spin or many-body system. 
(IV) In practice, speed control of a partial information of the system may be sufficient.
For example, we may want to control the speed of the center of mass of an atomic gas regardless of the motion of individual atoms. FFST can be classified according to whether it aims at speed control of the full or partial information of the model system.
(V) FFST can be classified according to whether the driving field is exact or approximate.
The driving field, which generates target states exactly, has been derived in several examples. 
On the other hand, it has been also shown, in many systems, that approximate driving fields, which are more experimentally feasible, can generate the target states with high fidelity. 
Although this robustness of the control is highly desirable, its origin and criteria for the robustness do not seem to be yet clear.
\begin{table}[!h]
\caption{Classification of fast-forward according to the nature of systems, objectives and driving field~\cite{Masuda_thesis}. (A row in the original table was merged with other one, and the order of the rows was modified from the original table.)}
\label{Classification of fast-forward}
{\renewcommand\arraystretch{1.2}
\begin{tabular}{llll}
\hline
(I) & classical  & quantum\\
(II) & relativistic & non-relativistic\\
(III) & single & many-body\\
(IV) &  partial &  total\\
(V) & exact & approximate\\
\hline
\end{tabular}
}
\label{table_Classification_3_4_22}
\end{table}

FFST also expanded in directions that were not anticipated when Table~\ref{table_Classification_3_4_22} was  made in 2008, such as protection of quantum state~\cite{Masuda2013}, ion sorting~\cite{Masuda2014}, generating an excited state from the ground state~\cite{Martinez-Garaot2016} and non-equilibrium statistical mechanics~\cite{Babajanova2018,Nakamura2020}. 
Problem of the singularity of the driving potential and lack of viable trajectories were mitigated~\cite{Patra2021,Masuda2021}. 
While most previous researches of FFST focused on acceleration, especially acceleration of adiabatic dynamics, FFST is also applicable to deceleration, stopping, and reversal of dynamics.
Recently, deceleration attracted attentions in terms of the mitigation of hardware limitations in quantum technologies~\cite{Masuda2021}.
It is expected that time-reversing and stopping dynamics will be studied more intensely elsewhere. 
A novel method of speed control called inter-trajectory travel (ITT) was derived from FFST. The ITT will find a wider application for example in other STA protocols. 



{\bf Author contributions}\\
SM authored mainly sections 1--4, 6 and 7, and KN authored mainly sections 5.

{\bf Acknowledgements}\\
S.M. acknowledges the support from JST [Moonshot R \& D] [Grant Number JPMJMS2061] and JSPS KAKENHI [Grant Number 18K03486].




\end{document}